\documentclass[
    prl, twocolumn, superscriptaddress, nofootinbib, amsmath, amssymb,
    aps, floatfix
]{revtex4-2}

\usepackage{graphicx}
\usepackage{svg}
\svgsetup{
    inkscapepath=i/svg-inkscape/
}
\svgpath{{svg/}}

\usepackage{xcolor}
\usepackage{setspace}
\usepackage{hyperref}

\usepackage[utf8]{inputenc}
\usepackage[T1]{fontenc}
\usepackage{lmodern}

\usepackage{booktabs}
\usepackage{color}
\usepackage{epsfig}
\usepackage{ifpdf}
\usepackage{amsmath}
\usepackage{bm}
\usepackage[english]{babel}
\usepackage{amsfonts}
\usepackage{amssymb}
\usepackage{braket}
\usepackage{enumerate}
\usepackage{cancel}
\usepackage{multirow}
\usepackage{cleveref}
\usepackage{xspace}
\usepackage{array}
\usepackage{fancyvrb}
\usepackage{fontawesome}
\usepackage{dcolumn}
\usepackage{slashed}
\usepackage[normalem]{ulem}
\usepackage{url}

\newcommand{\aap}{Astronomy \& Astrophysics}

\def\XXint#1#2#3{{\setbox0=\hbox{$#1{#2#3}{\int}$}
     \vcenter{\hbox{$#2#3$}}\kern-.5\wd0}}

\makeatletter
\g@addto@macro\bfseries{\boldmath}
\makeatother

\definecolor{nicered}{rgb}{0.7,0.1,0.1}
\definecolor{nicegreen}{rgb}{0.1,0.5,0.1}
\hypersetup{colorlinks, urlcolor=blue, citecolor=nicegreen,linkcolor= nicered}

\begin{document}
\title{Detecting Quadratically Coupled Ultra-light Dark Matter with Stimulated Annihilation}

\author{Yuanlin Gong}
\email{yuanlingong@nnu.edu.cn}
\affiliation{Department of Physics and Institute of Theoretical Physics, Nanjing Normal University, Nanjing, 210023, China}

\author{Xin Liu}
\email{liuxin@njnu.edu.cn}
\affiliation{Department of Physics and Institute of Theoretical Physics, Nanjing Normal University, Nanjing, 210023, China}

\author{Lei Wu}
\email{leiwu@njnu.edu.cn}
\affiliation{Department of Physics and Institute of Theoretical Physics, Nanjing Normal University, Nanjing, 210023, China}

\author{Qiaoli Yang}
\email{qiaoliyang@jnu.edu.cn}
\affiliation{Siyuan Laboratory and Department of Physics, Jinan University, Guangzhou 510632, China}

\author{Bin Zhu}
\email{zhubin@mail.nankai.edu.cn}
\affiliation{Department of Physics, Yantai University, Yantai 264005, China}

\begin{abstract}
Ultra-light Dark Matter (ULDM) is one of the most promising DM candidates. Due to the Bose enhancement, we find the annihilation rate of the ULDM in the presence of background photon radiation can be greatly enhanced and produce a distinctive reflected electromagnetic wave with an angular frequency equal to the ULDM mass. We propose to utilize such stimulated annihilation to probe the ULDM with the electromagnetic quadratic coupling by emitting a beam of radio into space. With a power of 50 MW emitter, we forecast the sensitivity of quadratic coupling in different local halo models for low-frequency radio telescopes, such as LOFAR, UTR-2 and ngLOBO.
\end{abstract}

\maketitle
\newpage

\section{Introduction}

\begin{figure*}[ht]
\centering
\includegraphics[width=0.5\linewidth]{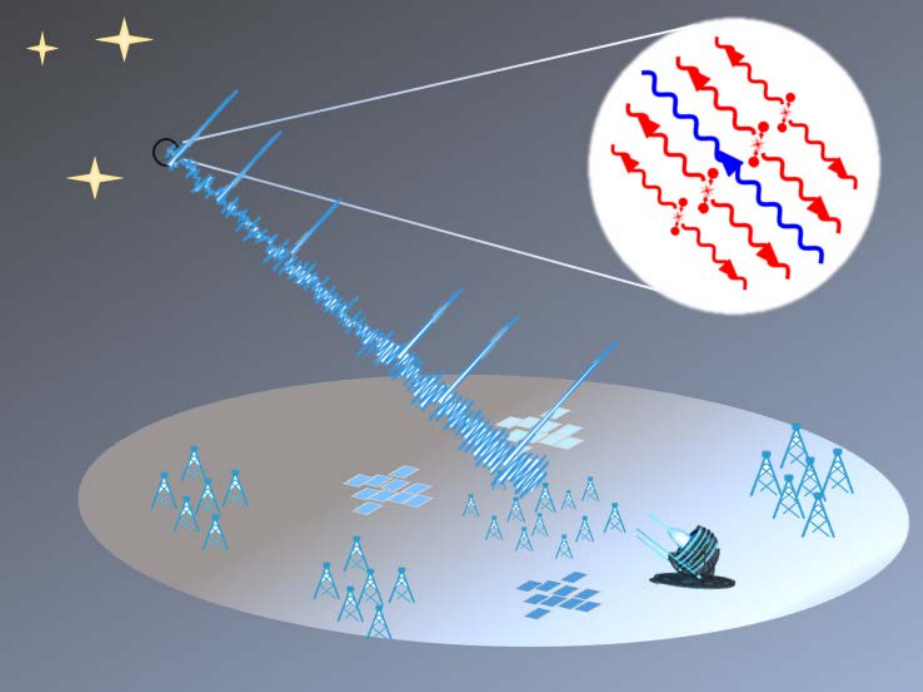}    \caption{Conceptual design of our proposed experiment. A powerful radio beam (blue wavy line) is sent to space to stimulate the annihilation of the ULDM (red bullet). The reflected radio (red wavy line) will be detected by the array telescope.}
\label{setup}
\end{figure*}

Dark matter, the invisible substance accounting for more than 80\% of the matter in the universe, continues to be a compelling mystery in modern physics~\cite{Bertone:2004pz}. The extensively studied canonical cold dark matter models, represented by weakly interacting massive particles, are attractive, providing the correct relic abundance, and simultaneously solving other modern puzzles, such as the hierarchy problem~\cite{Jungman:1995df}. However, the recent null experiment results put increasingly stringent constraints on various WIMP models~\cite{Roszkowski:2017nbc}.

The ultra-light dark matter (ULDM) that features a spin-0 particle with a mass ranging from $10^{-24}$~eV to eV is a competitive alternative, which could be produced through the misalignment mechanism or its variants in the early universe~\cite{Preskill:1982cy, Abbott:1982af, Dine:1982ah,Herring:2019hbe, Batell:2021ofv}. The wave-like nature of these particles can not only preserve the merits of cold dark matter but also may provide a solution to the small-scale structure problem~\cite{Hu:2000ke, Hui:2016ltb, Ferreira:2020fam}. Moreover, recent study on the galactic scale such as gravitationally lensed images indicates the increasing success of the wave-like DM versus the particle-like DM~\cite{AlfredAmruth:2023evq}. Besides, the ULDM also possibly addresses the Strong CP problem~\cite{Peccei:1977hh,Peccei:1977ur,Weinberg:1977ma,Wilczek:1977pj}, aforementioned hierarchy problem~\cite{Graham:2015cka}, and dark energy~\cite{Peccei:1987mm,Frieman:1995pm,Ferreira:2018wup}.

The distribution of the ULDM near Earth is crucial for its detection in laboratory experiments. Aside from the DM halo of the Milky Way halo~\cite{Turner:1985si, Navarro:1996gj, Pato:2015tja}, the caustic ring model describing the full phase space distribution of the Milky
Way is also well motivated and established \cite{Duffy:2008dk,Banik:2015vts,Chakrabarty:2020qgm}. It results in the occurrence of flows with varying density and dispersion near the Earth. On the other hand, the formation of a local ULDM dark halo~\cite{Banerjee:2019epw,Banerjee:2019xuy,Budker:2023sex}
bound to other gravitational sources, such as the Sun and Earth, provides an alternative profile possibility. The Sun and Earth halo result in much higher dark matter densities around the Earth's vicinity compared to the Milky Way halo over a wide mass range, which could significantly affect the sensitivity of ULDM detection.

Note that the approaches to detecting the ULDM strongly depend on its interactions with the Standard Model (SM) particles. Generically, the linear couplings of the ULDM with the SM particles are dominant, however, which have been tightly constrained by atom clocks ~\cite{Arvanitaki:2014faa, Derevianko:2013oaa,  Tsai:2021lly,Aharony:2019iad, Geraci:2018fax}, atomic spectroscopy~\cite{VanTilburg:2015oza, Stadnik:2015kia, Antypas:2019qji, Tretiak:2022ndx, Stadnik:2022gaa}, laser interferometry~\cite{Stadnik:2014tta,  Geraci:2016fva}, gravitational-wave detectors~\cite{Arvanitaki:2016fyj, Grote:2019uvn, Vermeulen:2021epa}, astrophysical probes~\cite{Olive:2007aj,Berlin:2016woy,  Khmelnitsky:2013lxt,Blas:2016ddr,Athron:2020maw,Yuan:2022nmu} and others~\cite{Savalle:2020vgz, Oswald:2021vtc,Ren:2021prq,Gu:2021lni,Su:2021jvk,Wang:2021uyb,Yang:2019xdz}. Besides, the enhanced emission of the axion clusters or axion-like particles may provide a complementary bound~\cite{Kephart:1986vc, Kephart:1994uy, Caputo:2018vmy}. On the other hand, the quadratic interactions of the ULDM dominating over the linear ones in recent years have drawn particular interest, such as in theories with $\mathbb{Z}_{N}$ symmetry~\cite{Hook:2018jle, DiLuzio:2021pxd} and relaxion mechanism~\cite{Banerjee:2022sqg}. Despite the above experimental bounds, the parameter space of the quadratic interactions remains largely unexplored.

In this article, we propose to probe the quadratic interaction of ULDM by detecting the reflected electromagnetic wave signal produced from the stimulated annihilation~(as shown in Fig.~\ref{setup}). In contrast with the spontaneous process, the flux of photons from the stimulated annihilation can be greatly enhanced by the presence of ambient photons, due to Bose enhancement. Since the power for signal photons highly depends on the local DM density, we consider four different local DM halos, the iso-thermal halo, the caustic ring model, the Sun halo and the Earth halo in our analysis. By using a radio emitter with a power of 50 MW and low-frequency radio telescopes, we present the new projected limit of the quadratic coupling of ULDM in the mass range of $2.07 \times 10^{-8}\mathrm{~eV} \sim 1 \times 10^{-6}\mathrm{~eV}$ for all of the halos.

\section{Stimulated Annihilation} 

The quadratic interaction of the ULDM field $\phi$ with the electromagnetic fields $A_\mu$ is given by,
\begin{equation}
\mathcal{L}=\frac{d_e^{(2)}}{4}\frac{2\pi}{M_{\mathrm{Pl}}^2} \phi^{2}{F}_{\mu \nu } F^{\mu \nu},
\label{Lagrangian}
\end{equation}
where the field strength $F^{ \mu \nu }=\partial ^\mu A^\nu -\partial^\nu A^\mu$, $d_e^{(2)}$ is the quadratic coupling constant, and $M_{\mathrm{Pl}}= 1.22\times10^{19}~\mathrm{GeV}$ is the Planck mass (for the pseudo-scalar case, see \cite{Arza:2019nta}). For simplicity, we assume $g'=\frac{2\pi}{M_{\mathrm{Pl}}^2}d_e^{(2)}$ in our calculations. 
Since the DM is non-relativistic in the halo, the angular frequency of the ULDM is approximate to its mass, $\omega_\phi=m_{\phi}$. With Eq.~\ref{Lagrangian}, we can have the cross-section of the spontaneous annihilation process $\phi\phi \to \gamma\gamma$,
\begin{equation}
\sigma_{0}=\frac{1}{32 \pi } \frac{1}{\beta} g'^2m_\phi^2,\label{total section}
\end{equation}
where the factor $\beta$ is the velocity of ULDM. Due to the tiny coupling and the small mass of the ULDM, the spontaneous annihilation rate is highly suppressed. We note that a certain incoming photon can stimulate the annihilation of the ULDM and thus produce an observable electromagnetic signal. There are two methodologies to describe the stimulated emission:  1) One involves Boltzmann equation in where the spontaneous annihilation amplitude of ULDM is enhanced by the Bose enhancement from the indistinguishability of identical bosons~\cite{Kephart:1994uy,Caputo:2018vmy,Buen-Abad:2021qvj}. 2) The one resorts to the modified electrodynamics, where the question become a radiation problem. The effective current density is provided by the interaction in Eq. \ref{Lagrangian}. The $\phi$ manifests itself as a classical coherent field and the stimulated enhancement is described through the resonance \cite{Arza:2019nta,Arza:2021nec}. We continue our proposal in method 1, while giving the derivation in method 2 in the Supplemental Material.

To obtain the production rate of the photons from stimulated annihilation, we first consider the number of states in the phase space of spontaneous annihilation process $\phi \phi \to \gamma \gamma$. In the 
vacuum, there are $f_{\phi}$ ULDMs with the certain momentum and zero photons  in the phase space of the initial states. After annihilation, the final state will contain $(f_{\phi}-1)$ ULDMs and two photons with two directions of momentum
that anti-parallel to each others in the center-of-mass frame of the two ULDMs, i.e.,  $|\mathrm{i}\rangle_0=|f_{\phi},f_{\phi};0,0\rangle, ~ |\mathrm{f}\rangle_0=|f_{\phi}-1,f_{\phi}-1;1,1\rangle.$ However, due to the Bose enhancement of identical photons, if the annihilation occurs in the background of $f_\gamma$ photons, this process will be stimulated and leads to
\begin{equation}
|\mathrm{i}\rangle=|f_{\phi},f_{\phi};f_{\gamma},f_{\gamma}\rangle,~ |\mathrm{f}\rangle=|f_{\phi}-1,f_{\phi}-1;f_{\gamma}+1,f_{\gamma}+1\rangle.
\end{equation}
Then, we can have the scattering amplitude of the stimulated annihilation
\begin{eqnarray}
    \mathcal{M}_{\mathrm{i\to f}}
    &=&\mathcal{M}_0^{\dagger}f_{\phi}(f_{\gamma}+1),
\end{eqnarray}
where $\mathcal{M}_{0}$ is the  spontaneous annihilation amplitude. The inverse process of two photons annihilating to two ULDMs in the vicinity of $f_\gamma$ ambient photons corresponds to the following matrix element:
\begin{equation}
    \mathcal{M}_{\mathrm{f\to i}}=\mathcal{M}_{0}(f_{\phi}+1)f_{\gamma}.
\end{equation}
In the presence of ambient photons, the effective annihilation amplitude for the stimulated annihilation of the ULDMs is determined by the difference between the amplitude of the annihilation and production processes of the photons:
\begin{eqnarray}
|\mathcal{M}_{\mathrm{i\to f}}|^{2}-|\mathcal{M}_{\mathrm{f\to i}}|^{2}
&=&|\mathcal{M}_0|^2[f_{\phi}^2(f_{\gamma}+1)^2-(f_{\phi}+1)^2f_{\gamma}^2],\nonumber\\
&=&|\mathcal{M}_0|^2[f_{\phi}^2+2f_{\gamma}f_{\phi}^2-2f_{\phi}f_{\gamma}^2-f_{\gamma}^2],\nonumber\\
&\approx&|\mathcal{M}_0|^2f_{\phi}^2(1+2f_{\gamma}).
\end{eqnarray}
The four terms in the second line are interpreted as the contributions from spontaneous annihilation, stimulated annihilation, inverse stimulated annihilation and inverse spontaneous annihilation. It is then clear that the factor $2f_\gamma$ is the enhancement for stimulated annihilation compared to spontaneous annihilation. We have used the approximation $f_{\phi}\gg f_{\gamma}$ in the last line.

From the Boltzmann equation, we can obtain the annihilation rate of the ULDMs from the stimulated annihilation process $\phi(p_1) \phi(p_2) \to \gamma(k_1) \gamma(k_2)$, 
\begin{eqnarray}
    \dot{n}_{\phi}
&=&-\int d\Pi_{\phi}d\Pi_{\phi}d\Pi_{\gamma}d\Pi_{\gamma}(2\pi)^{4}\delta^{4}(p_{1}+p_{2}-k_{1}-k_{2})\nonumber\\
&&\times\frac{1}{4}\cdot[|\mathcal{M}_{\mathrm{i\to f}}|^{2}-|\mathcal{M}_{\mathrm{f\to i}}|^{2}]\nonumber \\
&=&-4\beta n_{\phi}^2\sigma_{0}(1+2f_{\gamma}),
\label{varation of scalar}
\end{eqnarray}
where $d\Pi_{i}=g_{i}dp^{3}/((2\pi)^{3}\cdot2E_{i})$ is the usual phase-space volume. The factor of $1/4$ in the first line is the symmetry factor for identical particles in the initial and final states. $n_i$ is the number density of the ULDM or the photon, which is related to phase-space distribution by $n_i =\int\frac{g_i} {(2\pi)^3} f_i (\mathbf{p}) d^3p_i$. The production rate of the photons is the negative of the annihilation rate of the ULDMs.  This is greatly enhanced by the factor $2f_\gamma$, which arises from the stimulation of the ambient photons as $\omega_\gamma = m_\phi$. Moreover, the production rate of the photons depends on the ULDM density as $\rho_{\mathrm{DM}}^2$, rather than ${\rho_{\mathrm{DM}}}$.

A straightforward consideration inspired by the Bose enhancement and substantial occupation number of the ULDM is the more intriguing process of $n$ ULDMs scatter into two photons arising from the effective interactions described by the Lagrangian $\mathcal{L}=\frac{1}{4}\frac{d_e^{(n)}}{M_{\mathrm{Pl}}^{n}}\phi^n F_{\mu\nu}F^{\mu\nu}$, since 
 the annihilation rate of the ULDMs will dependent on $n_\phi^n$, being potential huge at the first glance. A closer derivation results that $\dot{n}_\phi$ is proportional to the factor $F=\left(\frac{\rho_\phi}{2M_{\mathrm{Pl}}^2m_\phi^2}\right)^n \frac{n^4}{n!}$. We find that this factor decreases in power with $n$ for the $m_\phi>1.07\times10^{-31}~\mathrm{eV}$ for all the halo model we considered here, despite the much higher density $\rho_\phi$ in the Earth and Sun halo at some mass ranges. For the  lower mass ranges , the ULDMs can not be the all dark matter and are constrained by the galaxy structure formation. Thus, as claimed before~\cite{Hook:2018jle, DiLuzio:2021pxd,Banerjee:2022sqg}, in the scenarios where the linear order $n=1$ is absent, the dominant contribution naturally emanates from the 
quadratic $n=2$ order, which is the focus of this work.

\section{Signal Power}

The signal power received by the telescope can be obtained by integrating the Eq.~\ref{varation of scalar} over time, the solid angle along the direction of the outgoing emission beam, the frequency, and the area
\begin{eqnarray}
    P&=&-\int dA d\nu d\Omega \int_0^{t_{\mathrm{off}}} dt  \dot n_\phi, \nonumber\\
    &=&\frac{1}{8}\frac{{g'}^2}{{m}_{\phi}^{2}}\frac{ P_0}{\Delta\nu_\phi }\int_0^{t_\mathrm{off}} {\rho_{\phi}^{2}}  {dt}.
\label{singal-power}
\end{eqnarray}
where the duration time of emission is denoted as $t_{\mathrm{off}}$. We ignored the interaction between the photon and the electron in the environment and factorized out the power of the source as,
\begin{eqnarray}
P_{\mathrm{0}}=\int dA d\nu d\Omega n_\gamma.
\end{eqnarray}
Here we assume that $f_\gamma$ is a Gaussian-like function with an expected value $\omega_\gamma$, and is related to 
$n_\gamma=\frac{2}{(2\pi)^3} 4\pi\omega_\gamma^{2}\Delta\omega_\gamma {f}_{\gamma}$ by averaging it over the bandwidth $\Delta\omega_\gamma=\Delta\omega_\phi\equiv 2\pi\Delta\nu_\phi$, with the ULDM bandwidth $\Delta\nu_\phi=2\nu_{\phi}\sigma_{v}$ depending on the velocity dispersion $\sigma_{v}$ of the ULDM~\cite{Caputo:2018vmy, Buen-Abad:2021qvj,Kephart:1994uy}. Note that $P_0$ is only determined by the properties of the emitter. In addition, it can be seen that the signal power $P$ in Eq.~\ref{singal-power} is sensitive to the local density of the ULDM halo, which is commonly assumed to be the iso-thermal DM halo. However, the distribution of ULDM in the Milky Way may differ from the predictions of the iso-thermal DM halo. The simulation for the formation of the Galactic halo suggests the enhancement (or suppression) of the local DM density due to the formation of “clumps” or streams \cite{Diemand:2008in}. Owing to the self-interactions or topological properties of the ULDM, the possibility of the formation of condensates \cite{Sikivie:2009qn}, clusters \cite{Kolb:1993zz}, boson stars \cite{JacksonKimball:2017qgk,Banerjee:2019epw, Banerjee:2019xuy,Budker:2023sex}, or domain walls ~\cite{Derevianko:2013oaa,Pospelov:2012mt} have been discussed . Their impact on the detection of ULDM was considered in \cite{Tsai:2021lly} for deep space atomic clocks, in \cite{Antypas:2019qji,Tretiak:2022ndx} for atomic spectroscopy, in \cite{Vermeulen:2021epa} for gravitational-wave detector and \cite{Gherghetta:2023myo} with neutrino oscillations. In our phenomenological study, we consider four different halo models,
\begin{itemize}
    \item Iso-thermal halo model. This standard density model of the Milky Way from N-body simulations predicts a local energy density of $\rho_{I} \approx 0.3 \mathrm{~GeV/cm^{3}}$ with velocity dispersion of $\sigma_{I}=270/\sqrt{3} \mathrm{~km/s}$~\cite{Turner:1985si, Navarro:1996gj, Pato:2015tja}. 
    \item Caustic ring model. Motivated by the description of the full phase space distribution of the Milky Way halo, the caustic ring model predicts the configuration of the caustics and flows \cite{Duffy:2008dk}. The properties of the flows are constrained by the observations of the IRAS and Gaia maps \cite{Chakrabarty:2020qgm}. The resulting local dark matter velocity distribution is dominated by the Big flow with velocity dispersion less than $\sigma_c=70$ m/s \cite{Banik:2015vts} and density as high as $\rho_c=10 \mathrm{GeV/cm^3}$ \cite{Chakrabarty:2020qgm}.
    \item Sun halo model. As an extension of the boson star, it has been recently discussed in \cite{Banerjee:2019epw, Banerjee:2019xuy,Tsai:2021lly,Budker:2023sex}. This was initially motivated by the suggestion from the large-scale numerical simulations of galaxy formation for fuzzy dark matter ($m_{\phi} \sim 10^{-22}~$eV) without \cite{Schive:2014dra, Veltmaat:2018dfz} and with \cite{Veltmaat:2019hou} the presence of baryons  that halo-like configuration can form in the central cores of galaxies. In the presence of quartic self-interactions and subsequent gravitational focusing, an external gravitational source, such as the Sun, in the background of virialized DM can effectively capture the ULDM, to form an over-density local halo. For the mass range of our interest($m > 10^{-9}$ eV, see below), its density $\rho_{\mathrm{sh}}$ around the Earth is much lower than that of the iso-thermal halo and decreases exponentially with the increase in the ULDM mass. The velocity dispersion is $1/(m_\phi R_{\mathrm{sh}})$, where $R_{\mathrm{sh}}$ is the radius of the Sun halo.  
    \item Earth halo model. Similar to the Sun halo model, it is also disscussed in \cite{Banerjee:2019epw, Banerjee:2019xuy,Budker:2023sex,Tsai:2021lly}.  \cite{Banerjee:2019epw} showed that the maximally allowed value of the energy density $\rho_{\mathrm{eh}}$ of the Earth halo is given by 
    \begin{equation}
    \rho_{\mathrm{eh}}(r) \propto\left\{\begin{array}{ll}
    \exp \left(-2 r / R_{\mathrm{eh}}\right) & \text { for } R_{\mathrm{eh}}>R_{\mathrm{E}}, \\
    \exp \left(-r^{2} / R_{\mathrm{eh}}^{2}\right) & \text { for } R_{\mathrm{eh}} \leq R_{\mathrm{E}}.
    \end{array}\right.\label{density}
    \end{equation}
    Here $R_{\mathrm{eh}}$ is the radius of the Earth halo, which is a function of the ULDM mass~\cite{Banerjee:2019epw}. $R_{\mathrm{E}}$ is the radius of the Earth. As a comparison with the iso-thermal DM halo, the Earth halo has a much higher density. On the other hand, with the increase in the ULDM mass, the Earth halo density decreases exponentially. For the estimation below, we take Supplementary Figure 2 in \cite{Banerjee:2019epw}, which provides a detailed description of the maximally allowed value of energy density $\rho_{\mathrm{eh}}$.
\end{itemize}

\section{Results and Discussions}

\begin{figure*}[t]
\centering
\includegraphics[width=0.6\linewidth]{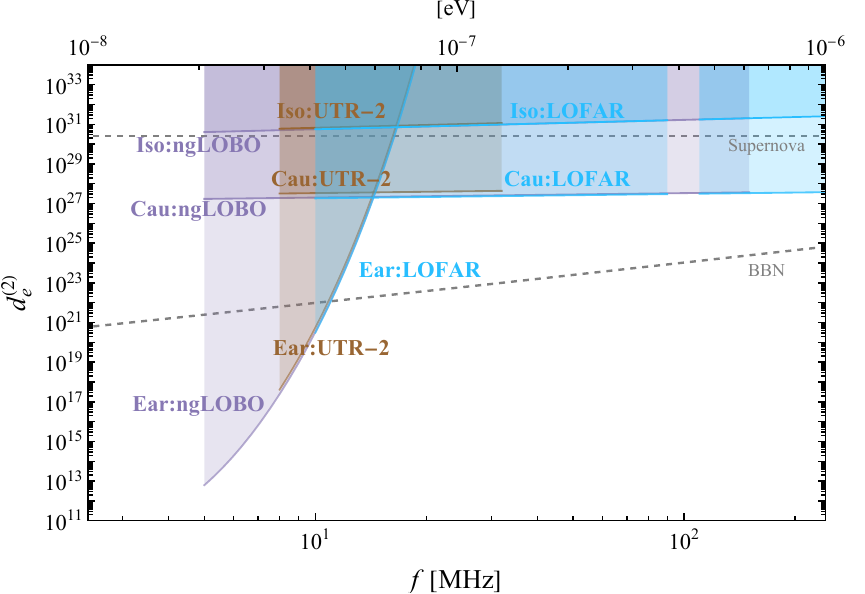}
\caption{The expected bounds on the plane of the dimensionless coupling $d_{e}^{(2)}$ versus the ULDM mass $m_\phi$ for the iso-thermal halo, caustic ring model and the Earth halo. The array telescopes LOFAR, UTR-2 and ngLOBO are considered here. The grey dashed lines are the limits  from the BBN and Supernova~\cite{Stadnik:2015kia}.} 
\label{projected limit}
\end{figure*}

In this work, our frequency range of interest is about $5 \sim 240\mathrm{MHz}$. For lower frequencies, the impact of the ionosphere will become significant, while for higher frequencies, the signal in will be suppressed by the mass of the ULDM as is seen in Eq. \ref{singal-power}. To detect such low-frequency radio signals, we use the LOFAR~\cite{2013A&A...556A...2V}, UTR-2~\cite{2016ExA....42...11K} and ngLOBO~\cite{2017arXiv170800090T} to estimate the sensitivity. LOFAR, as one of the new-generation radio telescopes, is capable of detecting radio signals in the frequency ranges $10~\mathrm{MHz}\sim 90 ~\mathrm{MHz}$, and $110\sim 240$ MHz with an unparalleled sensitivity owing to the novel phased-array design, dense core array, and long interferometric baselines ~\cite{2013A&A...556A...2V,2011A&A...530A..80S}.
Different from the traditional dish telescope, a number of dipole antenna elements are well arranged to compose a circular array with a diameter of $70\sim80$ meter, which enables it to observe in several modes and to detect the transient pulse signal with its high time and frequency resolution. The minimal frequency resolution of LOFAR is about 700 Hz \cite{2013A&A...556A...2V}. As for the UTR-2, a T-shaped antenna array can achieve a lower operating frequency range $8-32 ~\mathrm{MHz}$ with a frequency resolution down to 4 kHz \cite{2016ExA....42...11K}. The low band of the ngLOBO is to cover the frequency range $5-150~\mathrm{MHz}$~\cite{2017arXiv170800090T}, the resolution of which can reach 1 kHz at least ~\cite{5109716,2010amos.confE..59K}. The relevant parameters of the three telescopes are given in Table.~\ref{table1}.

\begin{table}
\begin{tabular}{cccc}
\hline\hline
\multicolumn{1}{c|}{Tele.}                 &\multicolumn{1}{c|}{ Fre.(MHz)}
&\multicolumn{1}{c|} {$T_{\mathrm{sys,m}}$(K)}
& $\Delta \nu_{\mathrm{res}}$(kHz)\\ \hline
\multicolumn{1}{c|}{LOFAR}
& \multicolumn{1}{c|}{10-90} 
& \multicolumn{1}{c|}{$3.5\times10^5$} & 0.7\\
\multicolumn{1}{c|}{LOFAR}
& \multicolumn{1}{c|}110-240
& \multicolumn{1}{c|}766 & 0.7\\
\multicolumn{1}{c|}{UTR-2}  
& \multicolumn{1}{c|}{8-32}  
& \multicolumn{1}{c|}{$6.1\times10^5$}& 4 \\
\multicolumn{1}{c|}{ngLOBO}
& \multicolumn{1}{c|}{5-150} 
& \multicolumn{1}{c|}{$2.0\times10^6$}&{$\leq$1}\\
\hline\hline
\end{tabular}
\caption{The frequency range,  maximum system temperature $T_{\mathrm{sys,m}}$
and frequency resolution $\Delta \nu_{\mathrm{res}}$ for LOFAR~\cite{2013A&A...556A...2V}, TUR-2~\cite{2016ExA....42...11K}, and ngLOBO \cite{2017arXiv170800090T}. The values of $T_{\mathrm{sys,m}}$ are evaluated at the frequencies $\nu=10 ~\mathrm{MHz}$, $\nu=110 ~\mathrm{MHz}$, $\nu=8 ~\mathrm{MHz}$, and $\nu=5 ~\mathrm{MHz}$. }
\label{table1}
\end{table}

The noise power of a radio array telescope with a frequency bandwidth $\Delta$ during the observing time $t_{\mathrm{off}}$ reads~\cite{ 2017isra.book.....T}
\begin{equation}
    P_{\mathrm{n}}=\frac{2 k T_\mathrm{sys}}{\eta_{\mathrm s}}\sqrt{\frac{\Delta}{n_{\mathrm{pol}}t_{\mathrm{off}}}},
\end{equation}
where $k$ is the Boltzmann constant and $n_{\mathrm{pol}}=2,1,2$ is the number of polarizations for LOFAR ~\cite{2013A&A...556A...2V}, UTR-2~\cite{2016ExA....42...11K} and ngLOBO~\cite{5109716,2010amos.confE..59K} respectively. For simplicity, we assume the detection efficiency $\eta_{s}=1$ in our numerical calculations. The bandwidth is determined by $\Delta=\mathrm{max}({\Delta\nu_\phi,\Delta\nu_{\mathrm{res}}})$, where $\Delta\nu_{\mathrm{res}}$ the telescope frequency resolution. $T_{\mathrm{sys}}$ is the temperature of the array system. It is caused by several inevitable noises, such as the cosmic microwave background, the environment, the galaxy, and the instrument. Among them, galaxy noise is dominant~\cite{1492591}. Then, we approximate the system temperature as $T_{\mathrm{sys}}\approx1.23\times 10^{8}~\mathrm{K}~({\mathrm{MHz}}/{\nu})^{2.55}$, where $\nu$ is the frequency of the noise photon~\cite{1492591,2013A&A...556A...2V}.

Besides, the transverse velocity $\Vec{v}_{\perp}$ of the ULDM perpendicular to the outgoing radio beam will result in the displacement of the reflected radio signal from the location of the outgoing power source. Thus, the duration time $t_{\mathrm{off}}$ in Eq.~\ref{singal-power} should be less than the effective time $t_{\mathrm{eff}}=C\frac{R}{\langle|\Vec{v}_{\perp}|\rangle}$, where $R$ is the radius of the array telescope. In the frequency range of our interest, $\langle |\Vec{v}_{\perp}| \rangle$ is about $124~\mathrm{km/s}$ for the iso-thermal halo~\cite{Arza:2019nta}, about 5 km/s for the caustic ring model \cite{Chakrabarty:2020qgm} and is about $ 1.2~\mathrm{km/s}$ for the Earth halo~\cite{Banerjee:2019epw}. For the Sun halo, the contribution for transverse velocity comes from velocity $v_{\mathrm{s}}=1.1~\mathrm{km/s}$ of ULDM in the frame of the Sun and the relative velocity $v_{\mathrm{r}}\approx 29~\mathrm{km/s}$ between the Sun and the Earth. Thus, we take ${\langle|\Vec{v}_{\perp}|\rangle}$ equals $30~\mathrm{km/s}$. In reality, the geometry factor $C$ depends on the specific configuration of the emitter relative to the collector in the experiment. Here we take $C=0.3$ to estimate the sensitivity. Since the height of the ionosphere is approximately 48 km, the time required for a radio wave pointing to the telescope to travel from the ionosphere to the telescope is 160 $\mu$s, which is greater than the effective time for the iso-thermal halo, and not much smaller than that of the caustic ring and the Sun halos. Meanwhile, we note that the density of the Earth halo will vanish before the radio reaches the ionosphere. Therefore, it is justifiable to disregard the interaction between photons and electrons in the environment.

For an emitter with power $P_{0}$, we can have the total energy consumption $E_{\mathrm{0}}=NP_{\mathrm{0}}t_{\mathrm{off}}$, where $N=m_{\phi}/2\Delta\nu_\phi$ is the emission times. In the following calculations, we set $E_{0}=10$ MWyear, $P_{0}=50$ MW and $R=50$ m. By requiring $P > P_{\mathrm{n}}$, we can obtain the exclusion limit of the dimensionless coupling $d^{(2)}_{e}$,
\begin{eqnarray}
d_{e}^{(2)}&<&D \cdot \left( \frac{\mathrm{50~\mathrm{MW}}}{{P}_{0}}   \frac{T_\mathrm{sys}}{3.5\times 10^5~ \mathrm{K}}\right)^{{\frac{1} {2}}}\cdot \left(\frac{\Delta}{n_\mathrm{plo}}\right)^{\frac{1}{4}}\label{limit}
\end{eqnarray}
with
\begin{equation}
D=\left\{\begin{array}{ll}
7.43\cdot 10^{28}\cdot\left( \frac{4219~\mathrm{sec}}{t_{\mathrm{off}} }\right)^{\frac{1}{4}}\left(\frac{m_\phi^{3}}{\mathrm{GeV}^{3}}\frac{\mathrm{GeV}^{7}}{\rho_{\mathrm{I}}^2 t_{\mathrm{eff}}}\right)^{\frac{1}{2}}, &\mathrm{Iso}\\
9.24\cdot 10^{27}\cdot\left( \frac{1~\mathrm{sec}}{t_{\mathrm{off}} }\right)^{\frac{1}{4}}\left(\frac{m_\phi^{3}}{\mathrm{GeV}^{3}}\frac{\mathrm{GeV}^{7}}{\rho_{\mathrm{c}}^2 t_{\mathrm{eff}}}\right)^{\frac{1}{2}}, &\mathrm{Cau}\\
1.80\cdot 10^{28}\cdot\left(\frac{14.7~\mathrm{sec}}{ t_{\mathrm{off}}}\right)^{\frac{1}{4}}\left(\frac{m_\phi^3}{\mathrm{GeV}^3}\frac{\mathrm{GeV}^7}{\int_0^{t_{\mathrm{eff}}}\rho_{\mathrm{sh}}^{2}dt}\right)^{\frac{1}{ 2}}, &\mathrm{Sun}\\
1.85\cdot 10^{28}\cdot\left(\frac{16~\mathrm{sec}}{ t_{\mathrm{off}}}\right)^{\frac{1}{4}}\left(\frac{m_\phi^3}{\mathrm{GeV}^3}\frac{\mathrm{GeV}^7}{\int_0^{t_{\mathrm{eff}}}\rho_{\mathrm{eh}}^{2}dt}\right)^{\frac{1}{ 2}}, &\mathrm{Ear}\\
\end{array}\right. \label{limit-1}  
\end{equation}
For the iso-thermal halo, we can take $\Delta=\Delta\nu_\phi$. It is because that the bandwidth of the ULDM is scaled as $15.09~\mathrm{kHz}\cdot (m/0.03 ~\mu \mathrm{eV})$, which is larger than the frequency resolution of the LOFAR, UTR-2 and ngLOBO. While for the caustic ring model, the Sun halo, and the Earth halo, since the frequency resolution $\Delta\nu_{\mathrm{res}}$ is larger than the  bandwidth $\Delta\nu_{\mathrm{\phi}}$ of ULDM in our mass range, we take $\Delta=0.7,~1,~4$ kHz in Eq.~\ref{limit} for LOFAR, UTR-2 and ngLOBO, respectively.

The resulting expected exclusion limits are shown in Fig.~\ref{projected limit}. We can see that the bounds for the quadratic coupling $d^{(2)}_e$ strongly depend on the ULDM halo models. They are around $d^{(2)}_e < 10^{27}$ and $10^{30}$ in the caustic ring model and the iso-thermal halo, respectively. The bounds for the iso-thermal halo are weaker than the limits from the Big Bang Nucleosynthesis (BBN) and Supernova~\cite{Stadnik:2015kia}, while that for the iso-thermal halo are weaker than the limit from BBN but stronger that the limit from Supernova by 2-3 orders of magnitude.  Since the density of Sun halo is very low in our mass range, the corresponding constraint is much weaker than others and thus is not shown. On the other hand, the limit on $d^{(2)}_e$ in the Earth halo model can be $10^{13}$ to $10^{34}$ in the range of $2.07 \times 10^{-8}\mathrm{~eV}< m_\phi < 8.27 \times 10^{-8}\mathrm{~eV}$ ($5 < f < 20$ MHz). The limits in this range can be at most 8 orders of magnitude stronger than that of the BBN and at most 17 orders of magnitude stronger than that of the Supernova. All the limits shown weaken with the increasing mass. Since the ULDM bandwidth $\Delta\nu_\phi$ for the caustic ring model is larger than the telescope frequency resolution $\Delta\nu_{\mathrm{res}}$ in the higher mass range, we should take $\Delta=\Delta\nu_{\phi}$ in these mass range and the limits on the caustic ring model will be further suppressed.  With the improved measurement of the Gaia maps \cite{ESA}, a more precise local DM distribution can be reached and reduce the uncertainty in the search for the DM in the future. Besides, we note that the sensitivity can be further enhanced by increasing the power of the emitter and enlarging the array telescope radius. In addition, a better frequency resolution of the telescope will improve the results in the same work frequency.

\section{Acknowledgement}
YG would like to thank Ariel Arza, Hyungjin Kim for useful discussions. This work is supported by the National Natural Science Foundation of China (NNSFC) under grants No. 12275134, No. 12335005, No. 12275232, No. 12147228, and No. 12150010. 

\bibliography{refs}

\onecolumngrid
\clearpage

\setcounter{page}{1}
\setcounter{equation}{0}
\setcounter{figure}{0}
\setcounter{table}{0}
\setcounter{section}{0}
\setcounter{subsection}{0}
\renewcommand{\theequation}{S.\arabic{equation}}
\renewcommand{\thefigure}{S\arabic{figure}}
\renewcommand{\thetable}{S\arabic{table}}
\renewcommand{\thesection}{\Roman{section}}
\renewcommand{\thesubsection}{\Alph{subsection}}

\newcommand{\ssection}[1]{
    \addtocounter{section}{1}
    \section{\thesection.~~~#1}
    \addtocounter{section}{-1}
    \refstepcounter{section}
}
\newcommand{\ssubsection}[1]{
    \addtocounter{subsection}{1}
    \subsection{\thesubsection.~~~#1}
    \addtocounter{subsection}{-1}
    \refstepcounter{subsection}
}
\newcommand{\fakeaffil}[2]{$^{#1}$\textit{#2}\\}

\thispagestyle{empty}
\begin{center}
    \begin{spacing}{1.2}
        \textbf{\large
            \hypertarget{sm}{Supplemental Material:} Detecting Quadratically Coupled Ultra-light Dark Matter with Stimulated Annihilation\\
        }
    \end{spacing}
    \par\smallskip
    Yuanlin Gong,$^{1}$
    Xin Liu,$^{1}$
    Lei Wu,$^{1}$
    Qiaoli Yang,$^{2}$
    and Bin Zhu,$^{3}$
    \par
    {\small
        \fakeaffil{1}{Department of Physics and Institute of Theoretical Physics, Nanjing Normal University, Nanjing, 210023, China}
       \fakeaffil{2}{Siyuan Laboratory, Physics Department, Jinan University, Guangzhou 510632, China}
        \fakeaffil{3}{Department of Physics, Yantai University, Yantai 264005, China}
        
    }

\end{center}
\par\smallskip
At low energy, the effective representation of the ULDM is that of a coherently oscillating classical field given by \cite{Foster:2017hbq} $\phi(\vec{x},t)=\phi_{0}\sum_j\alpha_j\sqrt{f(v_j)\Delta v}\cos(m_{\phi}t+\vec p_j\cdot \vec x+\phi_j)$, where the oscillating frequency equals the mass $m_{\phi}$ of the underlying particle, and $\phi_{0}=\sqrt{2 \rho_{\mathrm{DM}}}/m_{\phi}$, depending on the DM density $\rho_{\mathrm{DM}}$ and the mass, refers to the present-day oscillation amplitude, $\alpha_j$ is a random number of the Rayleigh distribution $P(\alpha_j)=\alpha_je^{-\alpha_j/2}$, $f(v_j)$  is the local dark matter speed distribution, $\phi_j$ is a phase factor and $\Delta v$ is the speed interval. Typically $f(v_j)$ is a narrow pick in most halo models, thus the ULDM can effectively be considered a mono-frequency field within the coherent time.

To derive the power of the reflected electromagnetic signal originating from the stimulated annihilation in the aspect of modified electrodynamics, we write again down the Lagrangian density of the Ultra-Light Dark Matter (ULDM) $\phi$ and electromagnetic fields $A_\mu$:
\begin{equation}
\begin{split}
\mathcal{L}=&-\frac{1}{4}F_{\mu \nu }F^{\mu \nu }-J^{\mu }A_{\mu }+\frac{1}{2}\left(\partial _{\mu }\partial ^{\mu }-m_\phi{}^2\right)\phi^2+\frac{1}{4}g^{\prime}\phi^{2}F_{\mu\nu}F^{\mu\nu}.
\end{split}
\end{equation}
where $F^{ \mu \nu  }=\partial ^\mu A^\nu -\partial^\nu A^\mu,J_\mu$ are electromagnetic tensor and current. $m_\phi,g'$ are the mass of ULDM, and the coupling between ULDM and electromagnetic field, respectively. The relevant equations of motion read
\begin{eqnarray}
    &&\partial _{\mu }F^{\mu \nu }-\frac{1}{2}g'\partial _{\mu }\left({\phi^{2} \epsilon }^{\nu \mu \lambda \tau }F_{\lambda \tau }\right)=J^{\nu },\label{nonlinear_equation}\\
    &&\left(\partial _{\mu }\partial ^{\mu }-m_\phi^2\right)\phi-2g'\phi\vec{E}\cdot \vec{B}=0.   
\end{eqnarray} 
For the free space, the existential Lagrangian is $-\frac{1}{4}F_{\mu \nu }F^{\mu \nu }$,which imply the below equation of motion
\begin{equation}
\partial _{\mu }F^{\mu \nu }=0.
\end{equation}
With the expression of $F^{\mu\nu}$ and $J^{\mu}$, we can translate the last three equations into the following ordinary formation
\begin{eqnarray}
&&\vec{\nabla }\cdot\vec{E}=\rho _{\text{el}}+g'\vec{\nabla}(\phi^{2}\vec{E}),\\
&&\vec{\nabla }\times \vec{B}-\partial _t \vec{E}=\vec{j}_{\text{el}}+g'\vec{\nabla }\times\left( \phi^{2}\vec{B}\right)+g'\partial_t \left(\phi^{2}\vec{E}\right),\label{modfied-equation1}\\
&&\vec{\nabla }\times \vec{E}+\partial _t\vec{B}=0,\\
&&\vec{\nabla }\cdot \vec{B}=0,
\end{eqnarray}
and
\begin{equation}
    \partial _t^2\phi-\nabla ^2\phi+m_{\phi}^2\phi=-g'\phi\vec{E}\cdot \vec{B},
\end{equation}
which may together be named as quadratically coupled ULDM electrodynamics. In the presence of the background electromagnetic field $\vec{E_0}(\Vec{x},t)$, $\vec{B_0}(\Vec{x},t)$, accounting for the similar status to the charge density and current density, we can define ULDM-induced charge density $\rho_{\phi}$ and current density $\vec{j}_\phi$ as
\begin{eqnarray}
    &&\rho _\phi=g'\vec{\nabla}(\phi^{2}\vec{E_0}),\\
    &&\vec{j}_\phi=g'\vec{\nabla }\times\left( \phi^{2}\vec{B_0}\right)+g'\partial_t \left(\phi^{2}\vec{E_0}\right). 
\end{eqnarray}
As we will see shortly, it was $\vec{j}_\phi$ that will radiate the reflected electromagnetic signal with the existence of the electromagnetic field $\vec{E_0}(\Vec{x},t)$, $\vec{B_0}(\Vec{x},t)$ that we send out to space to stimulated the annihilation of ULDM. 

Using radiation gauge $\vec{\nabla}\cdot \vec{A}=0$ and setting $\vec{j}_{\text{el}}=0$ in eq. \ref{modfied-equation1}, the ULDM-induced electromagnetic radiation $\vec{A}(\vec{x},t)$ is given by
\begin{equation}
   (\partial_{t}^2-\nabla^2)\vec{A}=g'\left[\phi^{2}\left(\vec{\nabla }\times \vec{B_0}\right)  +\partial_t \phi^{2}\vec{E_0}+\phi^{2}\partial_{t}\vec{E_0}\right]+\mathcal{O}(g'^2). \label{pertubation-radiation}
\end{equation}
 where we have neglected the term with $\vec{ \nabla }\phi$ due to the fact that $\vec{ \nabla }\phi \ll \partial_t\phi$.   
Usually, the solution to this inhomogeneous Helmholtz equation is a time-dependent function in terms of the retarded Green function. But we would like to focus on the resonance part, which will give us the backward electromagnetic wave relative to the incident electromagnetic wave we sent out. To this end, we expand $\vec{E}_0$ and $\vec{B}_0$ into Fourier modes 
\begin{eqnarray}
&&\vec{E}_0(\vec{x},t)=\Re\int d^3k\vec{E}_{0}(\vec{k})e^{i(\vec{k}\cdot\vec{x}-\omega t)},\label{fourier_expand1}\\
&&\vec{B}_0(\vec{x},t)=\Re\int d^3k\hat{k}\times\vec{E}_{0}(\vec{k})e^{i(\vec{k}\cdot\vec{x}-\omega t)}, 
\label{fourier_expand2}
\end{eqnarray}
with their inverse transformation given by
\begin{eqnarray}
    &&\vec{E}_{0}(\vec{k},t)=\Re\int d^3x\vec{E}_{0}(\vec{x},t)e^{-i\vec{k}\cdot\vec{x}},\\
    &&\hat{k}\times\vec{E}_{0}(\vec{k},t)=\Re\int d^3x \vec{B}_{0}(\vec{x},t)e^{-i\vec{k}\cdot\vec{x}}.
\end{eqnarray}
Also notes that $\omega=|\vec{k}|$. Substitute the eq. \ref{fourier_expand1} and \ref{fourier_expand2} into \ref{pertubation-radiation} resulting 
\begin{eqnarray}
    (\partial_{t}^2-\nabla^2)\vec{A}=&-g'\phi_{0}^{2}m_{\phi}\sin(2m_{\phi}t)\vec{E}_{0},\label{radiation_1} 
\end{eqnarray}
where the first and third terms in eq. \ref{pertubation-radiation} are canceled by each others due to the identity $\omega=|\vec{k}|$. We have also used $\phi(\vec{x},t)=\phi_{0}\cos(m_{\phi}t)$.
Then, in the spirit of our previous discussion, we therefore reasonably have the radiation exhibited as 
\begin{equation}
    \vec{A}(\vec{x},t)=\Re \int d^3k \vec{{A}}(\vec{k},t)e^{i\vec{k}\cdot
    \vec{x}}.
\end{equation}
Given this expression, one can proceed to perform the inverse transformation on the eq.(\ref{radiation_1}) and get
\begin{equation}
    \begin{array}{ll}
         &\left(\partial _t^2+\omega ^2\right)\vec{\mathcal{A}}(\vec{k},t)e^{i\omega t}=-{g'}\phi_{0}^{2}m_{\phi }{\sin}\left(2m_{\phi }t\right)\vec{E}_{0}(\vec{k})e^{-{i\omega t}},
         \end{array}
\end{equation}
where $\vec{\mathcal{A}}(\vec{k},t)$ is the amplitude of $\vec{A}(\vec{k},t)$ and assumed varies slowly in time with respect to $\vec{A}(\vec{k},t)$. It's obvious that the resonance for this radiation system occurs at $\omega- 2m_{\phi}=-\omega$, i.e. when $\omega=m_{\phi}$, the spontaneous annihilation of the ULDM into two photons is resonance enhanced when there are background electromagnetic fields with the same frequency as the photons produced from the ULDM. In term of $\vec{\mathcal{A}}(\vec{k},t)$, the last equation can further rewritten as:
\begin{equation}
\begin{array}{ll} \left[\left(\partial_t^2\vec{\mathcal{A}}+2{i\omega }\partial _t\vec{\mathcal{A}}+({i\omega })^2\vec{\mathcal{A}}\right)+\omega
^2\vec{\mathcal{A}}\right]e^{{i\omega t}}=-{ g'\phi_{0}^{2}}m_{\phi }{\sin}\left(2m_{\phi }t\right)\vec{E}_{0}(\vec{k})e^{-{i\omega t}}.
\end{array}
\end{equation}
To get the evolution with the time, we neglect $\partial_t^2\vec{\mathcal{A}}$ respect to $2{i\omega }\partial _t\vec{\mathcal{A}}$ and take solely the resonance term into consideration, thus giving rise to 
\begin{equation}
\partial_t\vec{\mathcal{A}}=\frac{1}{4\omega
}{g'\phi_{0}^{2}}m_{\phi }\vec{E}_{0}(\vec{k})e^{2i\left(m_{\phi }-\omega \right)t}.
\end{equation}
In the light of our assumption about the $\vec{\mathcal{A}}(\vec{k},t)$, we can solve the above equation by the initial condition $\vec{\mathcal{A}}(\Vec{k},0)=0$, and obtain the result displayed as 
\begin{equation}
    \vec{\mathcal{A}}\left(\vec{k},t\right)=\frac{1}{4\omega }{g'\phi_{0}^{2}}m_{\phi }\vec{E}_{0}\left(\vec{k}\right)e^{{i\epsilon
t}}\left(\frac{{\sin}({\epsilon t})}{\epsilon }\right),
\end{equation}
where $\epsilon=m_{\phi}-\omega$ is defined for convenience. With the approximation
\begin{equation}
    \left(\frac{\sin(\epsilon t)}{\epsilon}\right)^2 \to \pi t \delta(\epsilon),\label{limit-2}
\end{equation}
in large $t$, we work out the reflected signal power as \begin{equation}
         P=\frac{1}{16}{g'}^2\phi_{0}^4 m_{\phi}^2 t\int{d\omega }\frac{{d}P_0}{{d\omega }}(\omega )\pi \delta (\epsilon )=\frac{1}{8}\frac{{g'}^{2}\rho^{2}_{\mathrm{DM}}}{m_{\phi}^{2}}\frac{d P_{0}}{d\nu}t,
\end{equation}
where $P_0$ is the power of the background electromagnetic wave and is expressed as
\begin{equation}
    P_0=\int {d\omega } \frac{{d}P_0}{{d\omega }}(\omega ).
\end{equation}
We have also used the identity $\nu=\frac{\omega}{2\pi}=\frac{m_{\phi}}{2\pi}$ and $\phi_{0}=\sqrt{2 \rho_{\mathrm{DM}}}/m_{\phi}$. There are two things we would like to remind: 1) the signal power was obtained under the assumption of perfectly cold ULDM. 2) we did not consider the DM density profile. For the more completed treatment, we refer to further work as those done in \cite{Arza:2021nec}.

As a final note, with the similar calculation, it's not hard to figure out the signal power for the linear interaction $\mathcal{L}=\frac{1}{4}g\phi F_{\mu\nu}F^{\mu\nu}$ as
\begin{equation}
    P_1=\frac{1}{16}{g}^2\phi_{0}^2m_{\phi}^2 t\int{d\omega }\frac{{d}P_0}{{d\omega }}(\omega )\pi \delta (\frac{m_{\phi}}{2}-\omega )=\frac{1}{16}{{g}^{2}\rho}\frac{d P_{0}}{d\nu}t,
\end{equation}
where the spectral density $\frac{dP_0}{d\nu}$ is evaluated at $\nu=\frac{\omega}{2\pi}=\frac{m_{\phi}}{4\pi}$ this time. At the moment, it is clear from the delta function that the resonance takes place at $m_{\phi}=2\omega$, which is the stimulated decay of the ULDM similar to the axion case \cite{Arza:2019nta}. Also, we can expect the same signal expression as the axion case from the electromagnetic field duality.

\end{document}